# Artificial Kagome Spin Ice Phase Recognition from the Initial Magnetization Curve

Breno Cecchi, Nathan Cruz, Marcelo Knobel, and Kleber Roberto Pirota

"Gleb Wataghin" Institute of Physics, University of Campinas, 13083-859, Campinas, SP, Brazil

(Dated: November 3, 2022)

**Abstract**

Artificial spin ices (ASIs) are designable arrays of interacting nanomagnets that span a wide range of magnetic phases associated with a number of spin lattice models. Here, we demonstrate that the phase of an artificial kagome spin ice can be determined from its initial magnetization curve. As a proof of concept, micromagnetic simulations of these curves were performed starting from representative microstates of different phases of the system. We show that the curves are characterized by phase-specific features in such a way that a pattern recognition algorithm predicts the phase of the initial microstate with good reliability. This achievement represents a new strategy to identify phases in ASIs, easier and more accessible than magnetic imaging techniques normally used for this task.

## 1. INTRODUCTION

Artificial spin ices (ASIs) have proliferated over the past fifteen years [1], [2]. They are collections of interacting monodomain nanomagnets arranged in designable lattices defined by lithography. In turn, their behavior can be related to several spin models. The ability to directly probe their microstates and to tune the geometry and interactions of such systems has made it possible to study the associated statistical mechanics in an unprecedented way [3]. As a result, a wide range of phases have been theoretically predicted and/or experimentally observed. For example, ASIs can exhibit not only standard ferromagnetic [4]–[7] and antiferromagnetic [4], [6]–[9] phases, but also more exotic long range orderings related to the particular lattice geometry [10]–[16]. Even more interesting are the non-trivial, correlated disordered states they host due to frustration, such as the ones characteristic of ice-like models [17]–[20], spin liquids [21]–[23], spin glasses [24], [25] and Coulomb phases [22], [26]–[28].

The experimental investigation of the phase diagrams of ASIs is mostly done through magnetic imaging techniques, such as magnetic force microscopy and photoemission electron microscopy combined with X-ray magnetic circular dichroism [3]. They enable the visualization of each magnetic moment's orientation in real space and time. This makes it possible to know the precise microstates that the system accessed, which is normally impossible when working with bulk materials. In theory, one has all the knowledge required to characterize the occurring phases, phase transitions and kinetics with this information. For example, a common experiment involves firstly demagnetizing the sample through a magnetic field protocol to bring the array of nanomagnets to a certain frozen configuration. Then, taking the sample to the proper microscope, it is possible to identify the phase and even the effective temperature of the system by comparing properties extractable from images, such as magnetic moment correlations and vertex populations, against predictions of spin models [13], [21], [29].

However, there are some practical drawbacks to this image-based strategy. Good high-contrast images can take hours or even days to be acquired (taking into account procedures like sample preparation, microscopy calibration, and image processing). Additionally, some experiments can only be carried out in a small number of particular locations, such as synchrotron light sources. Given this, it makes sense to have quicker and easier ways to ascertain the thermodynamic characteristics of ASIs.

In this work, we propose a strategy to identify the phase of an ASI from its initial magnetization curve, which can be readily measured by several standard magnetometry techniques. Despite their versatility and easy handling, magnetization curves have not been used for this task. The main reason is that the measured magnetization is a property of the whole sample, making it extremely difficult to infer any microstate.

In the following, we concentrate on artificial kagome spin ice (AKSI), made up of nanomagnets arranged on a kagome lattice, because the corresponding spin model is known to have a particularly rich phase diagram with four phases [30]–[32]. At high temperatures, the system is in a paramagnetic (PM) phase, characterized by an uncorrelated disorder. By lowering the temperature, it smoothly accesses a spin liquid phase referred to as spin ice 1 (SI1). In this regime, the system is still disordered at large length scales but locally obeys the kagome ice rules. Namely, they dictate each vertex has a two-in/one-out or one-in/two-out spin configuration. As the system cools down, it undergoes a first phase transition into the intriguing spin ice 2 (SI2) phase, a new spin liquid phase with a number of novel properties. Quite interestingly, here the spins fluctuate only through collective loop moves, preserving the ice rules but also giving rise to a long-range order (LRO) of magnetic charges. The spin state exhibits features of a Coulomb phase and can be seen as the coexistence of order and disorder as a result of the spin fragmentation phenomenon [3], [27]. At a still lower temperature, the system finally experiences a second phase transition and reaches its sixfold degenerate ground state (GS) with LRO of both spins and charges.

The primary hypothesis of this work is that the initial magnetization curve of an AKSI retains characteristics related to its starting point in such a way that the phase of the initial microstate can be deduced solely from the curve. If one applies a magnetic field to drive an AKSI from a given microstate to saturation, the exact way the system evolves depends on the initial microstate. On the other hand, microstates of the same phase obey characteristic constraints, as described in the last paragraph. Thus, it is reasonable to expect that evolutions initiating at microstates of the same phase share more resemblance than evolutions initiating at microstates of different phases. In turn, the magnetization curves of these processes should also present phase-specific features.

As a proof of concept, we ran micromagnetic simulations of initial magnetization curves of AKSI starting from various microstates, each one representative of a certain phase. Using some of these curves as a training set, we developed a pattern recognition algorithm that was able to identify the associated phase of a given curve of the remaining set with a fairly good performance. This accomplishment paves the way for novel methods of phase recognition in ASIs, and we anticipate that it will spur additional experimental work.

## 2. METHODOLOGY

A proper description of the thermodynamic behavior of AKSI must take into account the dipolar interaction between the nanomagnets [3], [21], [33]. Thus, we consider the dipolar hamiltonian

$$H = -\sum_{i \neq j} J_{ij} S_i S_j$$

with

$$J_{ij} = -\frac{\mu_0 m^2}{4\pi} \frac{\hat{e}_i \cdot \hat{e}_j - 3(\hat{e}_i \cdot \hat{r}_{ij})(\hat{e}_j \cdot \hat{r}_{ij})}{r_{ij}^3},$$

where $m\vec{S}_k$ is a magnetic dipole of magnitude $m$, located at $\vec{r}_k$, associated with an Ising spin $\vec{S}_k = S_k \hat{e}_k$ allowed to point only along direction $\hat{e}_k$ with $S_k = \pm 1$. A pair of spins $\vec{S}_i$ and $\vec{S}_j$, separated by $\vec{r}_{ij} = \vec{r}_j - \vec{r}_i$, has the corresponding coupling strength $J_{ij}$.

From this dipolar model, we sampled 20 distinct microstates for five temperatures of each of the four phases of the system. This amounts to four sets of 100 microstates, each set representing a given phase. The microstates were randomly generated by performing Monte Carlo simulations of a kagome lattice with $12 \times 12 \times 3$ spins with periodic boundary conditions. We employed the Metropolis algorithm with both single spin flip and spin loop flip dynamics and used 200 Monte Carlo steps per spin (MCS/spin) for thermalization. The temperatures were chosen to be well spaced, but also as distant as possible from the transition temperatures. The latter were estimated from the peaks of the specific heat, shown in the Supplemental Material, and were found to be $k_B T/|J_1| \sim 2.0$ for PM/SI1, $\sim 0.17$ for SI1/SI2 and $\sim 0.089$ for SI2/LRO, where $J_1$ is the nearest neighbors coupling strength.

We emphasize that the sets of microstates associated with the PM, SI1 and SI2 phases are representative samples of these extensively degenerate phases. Indeed, the spin correlations of the selected microstates distribute well around the mean values of the spin correlations of the dipolar kagome spin ice model, as shown in Fig. 1. These microstates were sampled at each 10 MCS/spin, which is longer than the autocorrelation time of the system.

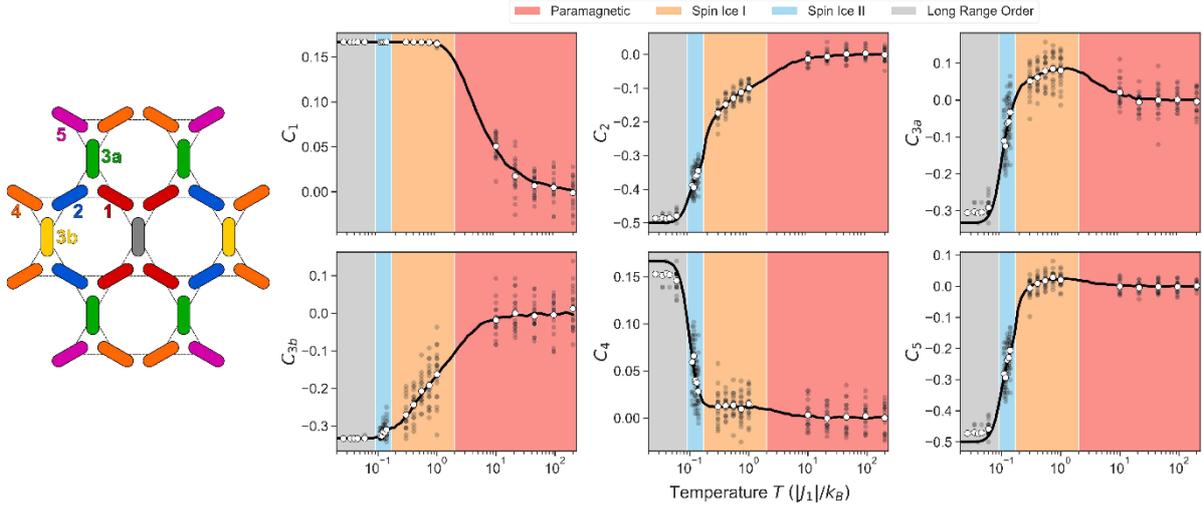

FIG. 1: Spin correlations $C_n = <\vec{S}_i \cdot \vec{S}_j>_n$ between $n$-th neighbors of kagome spin ice for $n = 1, 2, 3a, 3b, 4, 5$ (as indicated in the left panel) as a function of temperature $T$. Solid black lines are the thermodynamic mean values of the dipolar model. Gray dots are the values for each sampled microstate and the white dots are the averages over all microstates sampled at the same temperature.

On the other hand, the long-range ordered GS is only sixfold degenerate. However, as explained later, our pattern recognition algorithm needs a training set with more than just six datasets for being able to identify a given phase. In order to include the LRO phase in our analysis, the microstates sampled in the GS temperature range are actually a mixture, due to fluctuations, of LRO and SI2 phases but with considerable predominance of LRO. Because of this, we will refer to this phase as "partial long-range order" (PLRO). Namely, 84% to 89% of the spins obey the LRO pattern and the remaining part respects only SI2 constraints (see the LRO fraction distribution in Supplemental Material). As a result, the spin correlations distribution is centered slightly off the GS values but are still markedly different from the ones of the other phases, as seen in Fig. 1. To set up this set, Monte Carlo simulations were run for a long enough time to generate 20 distinct microstates for each temperature (although fluctuations become increasingly rare at low temperatures, they still eventually occur).

The whole set of 400 distinct microstates were used as the initial states in micromagnetic simulations of an analogous kagome lattice with $12 \times 12 \times 3$ nanomagnets. In order to do it, we set the magnetization of each nanomagnet to be uniform with the same orientation of the corresponding spin of the microstate, as illustrated in Fig. 2. The system was relaxed before applying the field in the y direction (Fig. 2) from zero to a high saturating value. In these simulations, we considered open boundary conditions, as this is the case in a real sample, and stadium shaped nanomagnets with length of 300 nm, width of 100 nm and thickness of 20 nm. The micromagnetic simulations were performed with Mumax3 [34] using cell dimensions of $4.4$ nm $\times$ $4.2$ nm $\times$ $20$ nm and material parameters of permalloy: saturation magnetization $M_s = 8 \times 10^5$ A/m and exchange stiffness $A_{ex} = 1 \times 10^{-11}$ J/m.

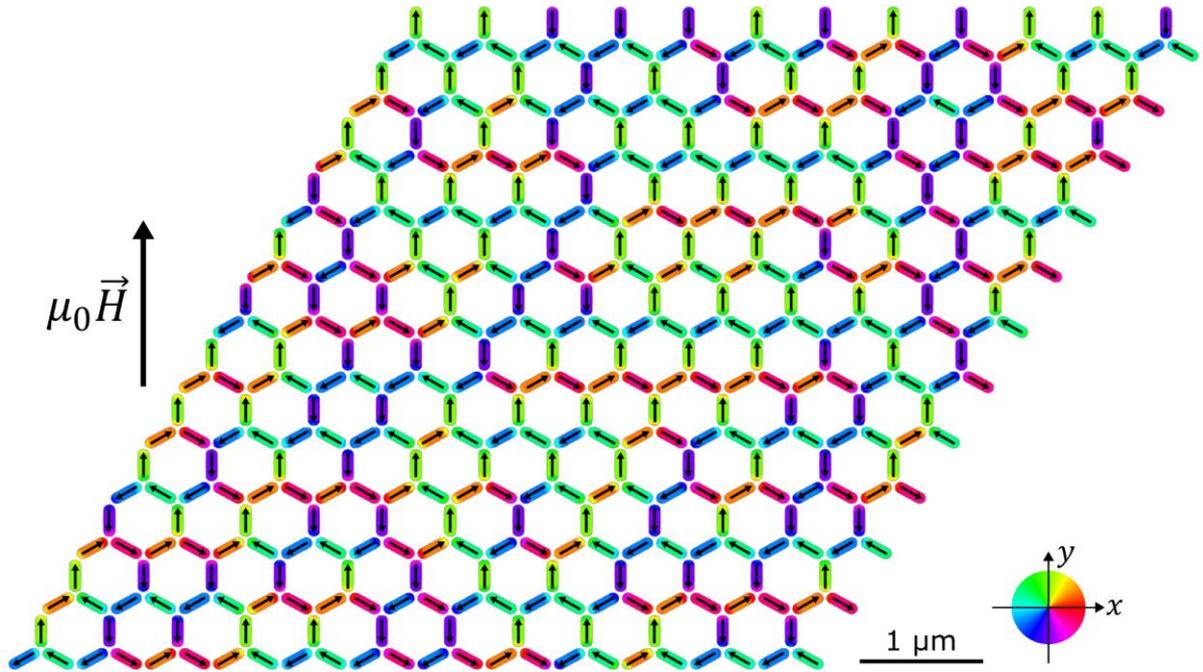

FIG. 2: An example of an initial state in micromagnetic simulations. The arrows indicate the spins of a microstate representative of the SI1 phase, taken from Monte Carlo simulations. To set the initial state, an uniform magnetization was assigned to each nanomagnet following the same orientation of the corresponding spin and the system was relaxed. The initial magnetization curve was simulated by applying a magnetic field $\mu_0 \vec{H}$ in the y direction. The color code represents the local direction of the magnetization.

## 3. RESULTS AND DISCUSSION

We defined common parameters for all 400 initial magnetization curves in order to characterize them. Each curve was divided into three regions separated by the boundary fields $\mu_0 H_{B1}$ and $\mu_0 H_{B2}$, as shown in Fig. 3(a). We call "lag interval" (Fig. 3(b)) the first part of the magnetization curve, for $0 \leq H \leq H_{B1}$, where the behavior is essentially linear. In this region, the curve is well fitted by a straight line, whose angular coefficient can be identified as the initial magnetic susceptibility $\chi_i$.

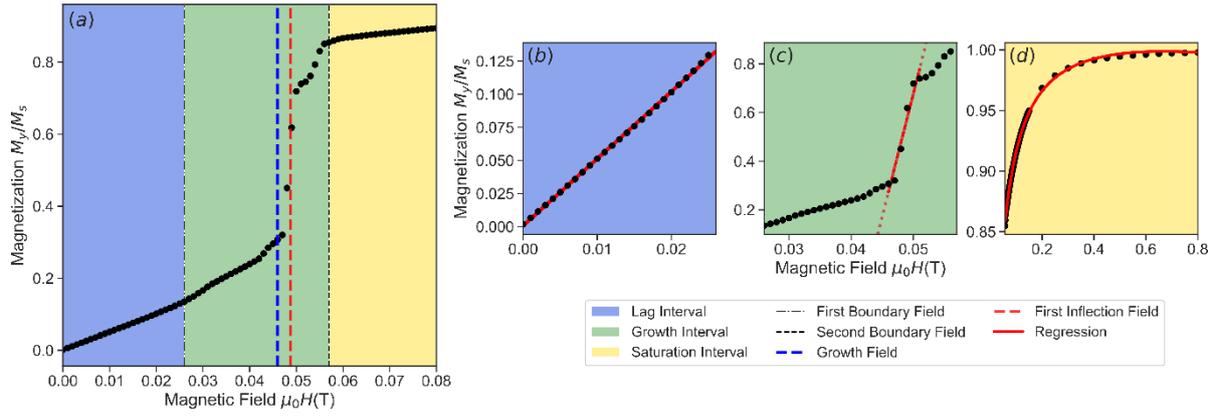

FIG. 3. (a) Initial magnetization curve starting from a SI2 microstate. It shows the three intervals separated by the boundary fields $\mu_0 H_{B1}$ and $\mu_0 H_{B2}$. (b) Lag interval, where we perform a linear fit to extract the initial magnetic susceptibility $\chi_i$ as the angular coefficient. (c) Growth interval, where we calculate the perimeter $p_G$ of the curve, the growth field $\mu_0 H_G$, the first inflection field $\mu_0 H_{if}$ and the angular coefficient $\alpha_G$ of the linear fit in the magnetization's steep rise region. (d) Saturation interval, where magnetization follows the LAS.

The "growth interval" (Fig. 3(c)), corresponding to $H_{B1} \leq H \leq H_{B2}$, begins at $H = H_{B1}$, where the curve starts deviating from its initial linear trend, and ends at $H = H_{B2}$, where the sample starts approaching saturation (see Supplemental Material for the precise procedure used to identify the values of $\mu_0 H_{B1}$ and $\mu_0 H_{B2}$). In this interval, the curve's behavior is more complex, containing most of its particular features. Here, we defined four parameters: the growth perimeter $p_G$ of the curve; the "growth field" $\mu_0 H_G$, taken as the field at which occurs a sudden change in the curve's derivative (see Supplemental Material); the "first inflection field" $\mu_0 H_{if}$, defined as the field value of the first inflection point; and the angular coefficient $\alpha_G$ of the linear fit performed in the subregion of steep increase of magnetization.

Finally, in the "saturation interval" (Fig. 3(d)), where $H \geq H_{B2}$, the magnetization follows the well known law of approach to saturation (LAS) [35]. However, this interval does not carry any information about the initial configuration since the magnetic moments of all nanomagnets have essentially aligned with the field.

The following six parameters have proved useful in identifying the AKSI phases: $\mu_0 H_{B1}, \chi_i, p_G, \mu_0 H_G, \mu_0 H_{if}$ and $\alpha_G$. The first evidence of that is shown in Fig. 4, which displays their distributions for each phase. Note how the PM and SI1 distributions of most parameters are well separated from one another and from the ones of SI2 and PLRO. This is particularly clear in the $\mu_0 H_G$ distribution (Fig. 4(d)). However, for all cases, the SI2 and PLRO distributions are centered at very close values and show considerable overlap.

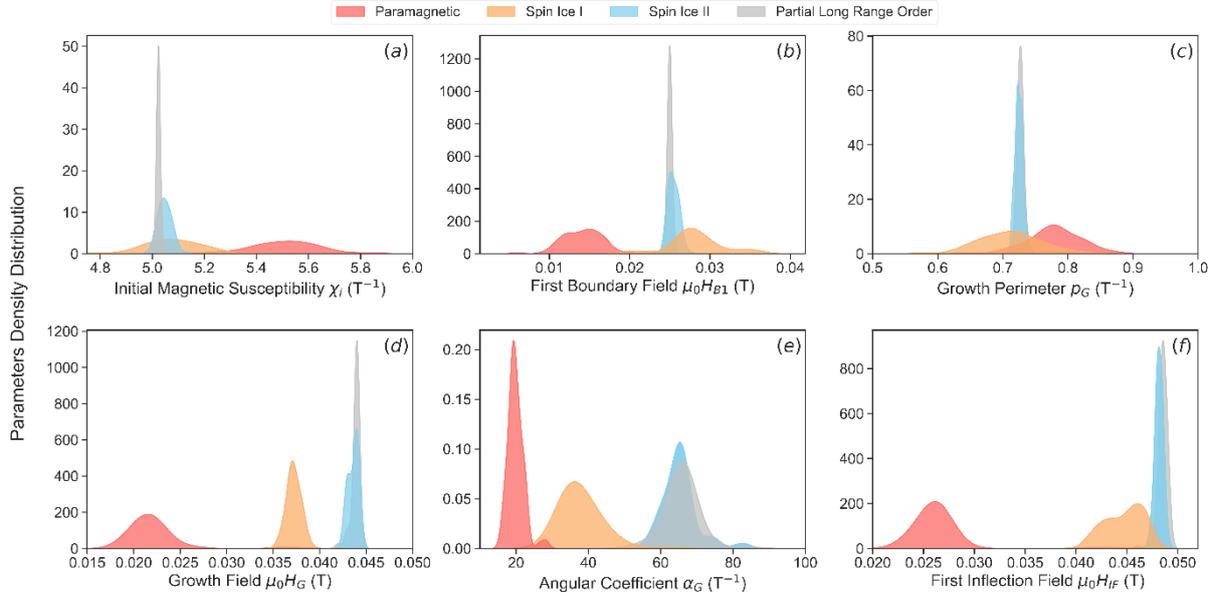

FIG. 4. Normalized distributions, for each phase, of the (a) initial magnetic susceptibility $\chi_i$, (b) first boundary field $\mu_0 H_{B1}$, (c) growth perimeter $p_G$, (d) growth field $\mu_0 H_G$, (e) angular coefficient $\alpha_G$, and (f) first inflection field $\mu_0 H_{if}$.

Another way to visualize the parameters' distribution comes from applying a principal component analysis (PCA) to our data. In this method, the parameters describe a six-dimensional space, where each of our curves is represented by a point. In this space, the vector for which the variance of the projected points is maximized is called the first principal component; and its orthonormal vector that maximizes the variance is called the second principal component. The two-dimensional subspace spanned by the first and second principal components, shown in Fig. 5, is the plane that better represents the data distribution. This dimensionality reduction retains meaningful features of the original data and allows one to visualize clusters of points sharing common properties. Indeed, one sees again that the PM, SI1 and SI2∪PLRO sets of points are well separated, whereas the SI2 and PLRO sets overlap. It is also worth noting that the greater the phase degeneracy the wider its parameters' distribution and, consequently, the larger its cluster.

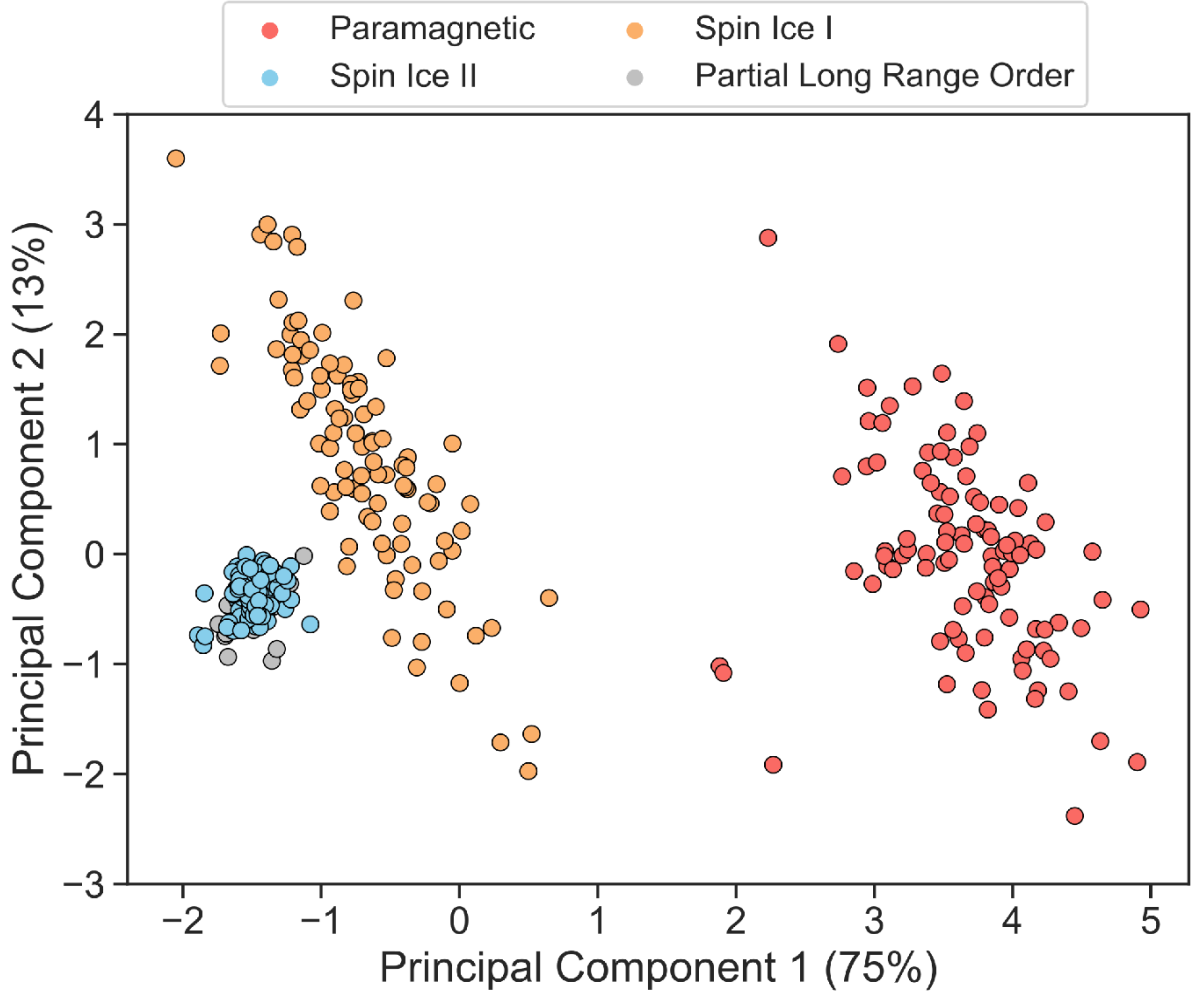

FIG. 5. PCA scatter plot of the points representing the six parameters of each initial magnetization curve. The data variance with respect to the first and second principal components are, respectively, 75% and 13% of the total variance.

The parameters' distributions shown in Figs. 4 and 5 strongly suggest that it is possible to identify the associated phase of each curve, with the exception of distinguishing between SI2 and PLRO. To further verify it, 70% of the curves were used as a training dataset for a supervised classification algorithm, the so-called support vector machine [36], to classify the remaining 30% of the test curves. The performance is summarized by the confusion matrix, shown in Fig. 6. The algorithm correctly predicted the phase of all 35 PM and 36 SI1 test curves. In addition, it correctly determined the phase of 23 out of the 28 SI2 test curves, where the other 5 were misclassified as PLRO; and it correctly determined the phase of 24 out of the 31 PLRO curves, where the other 7 were misclassified as SI2.

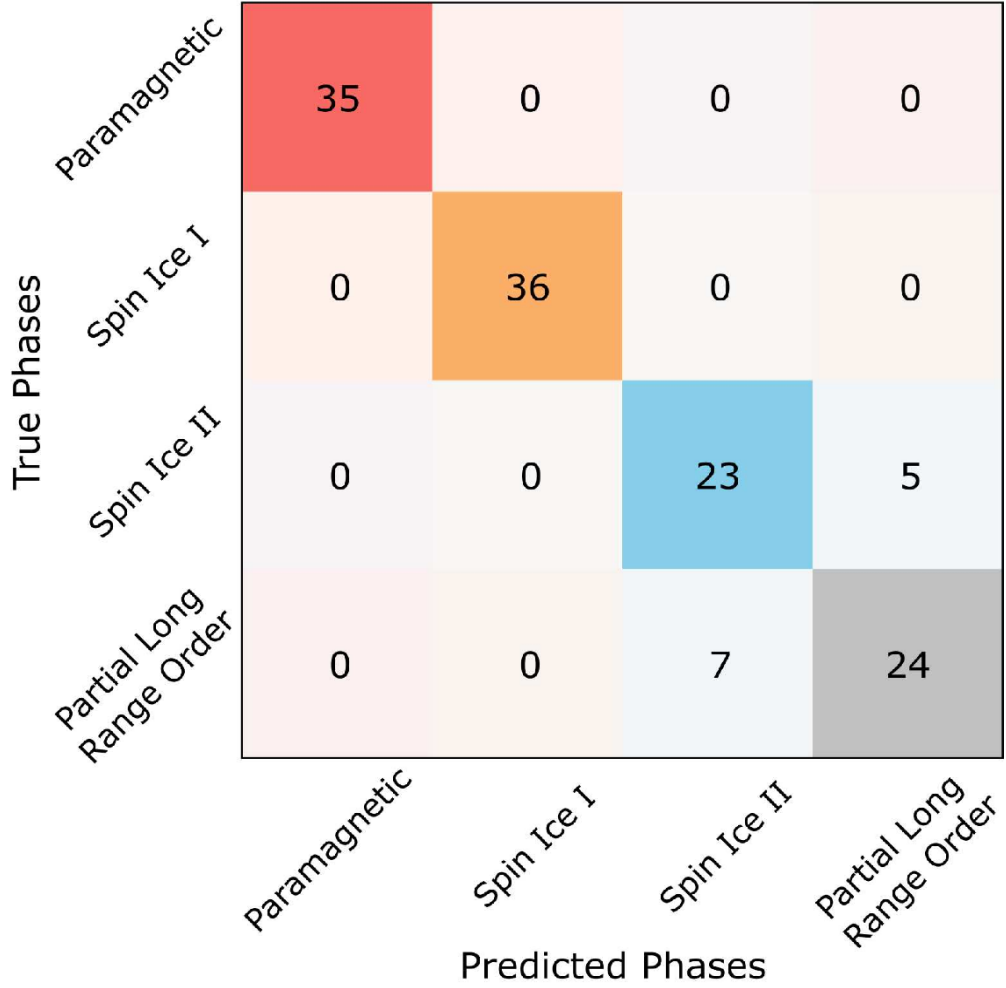

FIG. 6. Confusion matrix of our supervised classification algorithm developed to predict the associated phase of a given initial magnetization curve.

We consider the general performance of the algorithm fairly good. As expected, it showed maximum efficiency in recognizing the PM and SI1 phases and made really few mistakes in predictions regarding the SI2 and PLRO phases. Even so, the distinction between SI2 and PLRO is astonishingly good, considering the substantial overlap of their parameters' distributions. Additionally, part of the errors may be related to the mix of LRO and SI2 in the PLRO microstates.

Finally, we address some of the physics underlying our six parameters. Fig 7 shows how they change as a function of the effective temperature. Even when not monotonically, one sees that, as the temperature increases, $\chi_i$ increases (Fig. 7(a)) and $\mu_0 H_{B1}$, $\mu_0 H_{if}$ and $\mu_0 H_G$ decreases (Figs 7(b), 7(f) and 7(d), respectively). All these general trends indicate the magnetic behavior of AKSI gets softer with increasing temperature. We also observe that $\mu_0 H_{B2}$ remains practically constant, as shown in the Supplemental Material, which is reasonable since the saturation field should be the same for demagnetized initial states. In consequence of the magnetic softening with temperature and of the constancy of $\mu_0 H_{B2}$, $\alpha_G$ decreases and $p_G$ increases with temperature, as can be seen in Figs. 7(e) and 7(c). As a matter of fact, some of these parameters can be interesting indicators to verify phase transitions in complex engineered systems.

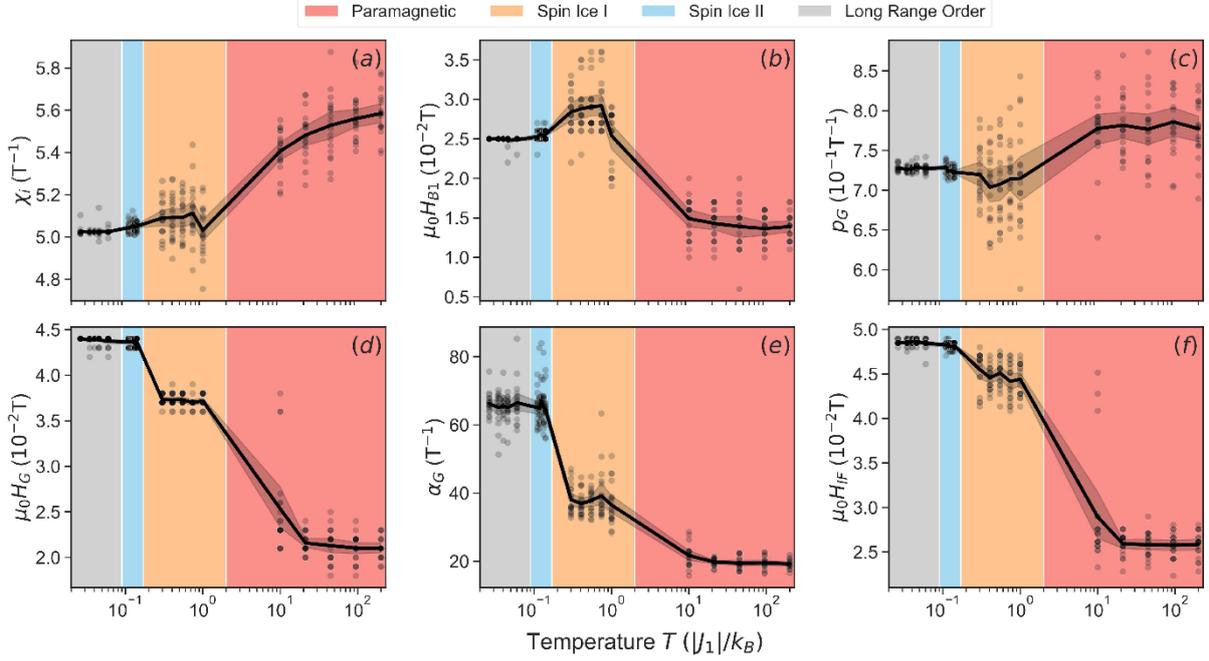

FIG. 7. Dependence on temperature of the (a) initial magnetic susceptibility $\chi_i$, (b) first boundary field $\mu_0 H_{B1}$, (c) growth perimeter $p_G$, (d) growth field $\mu_0 H_G$, (e) angular coefficient $\alpha_G$, and (f) first inflection field $\mu_0 H_{if}$. Gray dots are the parameters' values for each sampled microstate and the black curve is the average over all microstates sampled at the same temperature.

## 4. CONCLUSIONS AND PERSPECTIVES

This study's key finding is that an AKSI's initial magnetization curve preserves a significant amount of information on the phase of its original microstate. We showed that such curves may be utilized to reliably identify the phases using a pattern recognition algorithm. However, more research with bigger samples is needed to assess how the separation between the SI2 and PLRO phases improves when the SI2 proportion in the mixed SI2/LRO microstates decreases. In order to better understand the parameter distribution and see how they react at the critical points, it would also be intriguing to include more temperatures that are closer to the phase transitions.

Additionally, it will be rather simple to apply similar research to different ASI geometries. Our findings thus open the door to the experimental determination of ASI magnetic phases using only magnetometry techniques, which are more widely utilized and user-friendly than the magnetic imaging methods often employed for this type of study. Moreover, the phase recognition could be carried out in the exact same magnetometer right after a demagnetization protocol. By examining the patterns shared by a certain collection of magnetization curves that are challenging to recognize by human eye inspection, one might be able to decode the physics inherent in a particular ASI magnetization process. We are confident that, once improved, this will grow into a potent analysis method for identifying complex magnetic phases, possibly expanding future experimental studies.

## ACKNOWLEDGMENTS

The authors acknowledge David Navas for help on writing the micromagnetic simulations code. This research was financially supported by São Paulo Research Foundation through grants 2017/10581-1 and 2019/23317-6.

# SUPPLEMENTAL MATERIAL

# Artificial Kagome Spin Ice Phase Recognition from the Initial Magnetization Curve

Breno Cecchi, Nathan Cruz, Marcelo Knobel, and Kleber Roberto Pirota

"Gleb Wataghin" Institute of Physics, University of Campinas, 13083-859, Campinas, SP, Brazil

(Dated: November 3, 2022)

## 1. TRANSITION TEMPERATURES

The dipolar kagome spin ice (DKSI) model is known to have four phases [1]–[3]: paramagnetic (PM), spin ice 1 (SI1), spin ice 2 (SI2) and long-range order (LRO). We estimated the transition temperatures of the model from the peaks of the curve of specific heat $c$ as a function of temperature $T$, shown in Fig. S1. This was calculated by Monte Carlo simulations of a $12 \times 12$ kagome lattice with periodic boundary conditions, using 200 Monte Carlo steps per spin for thermalization and more $\sim 1000$ for measurements. The maxima occur at $k_B T/|J_1| = 2.0$, $0.17$ and $0.089$, corresponding to the PM/SI1, SI1/SI2 and SI2/LRO transitions, respectively, where $J_1$ is the nearest neighbor coupling strength.

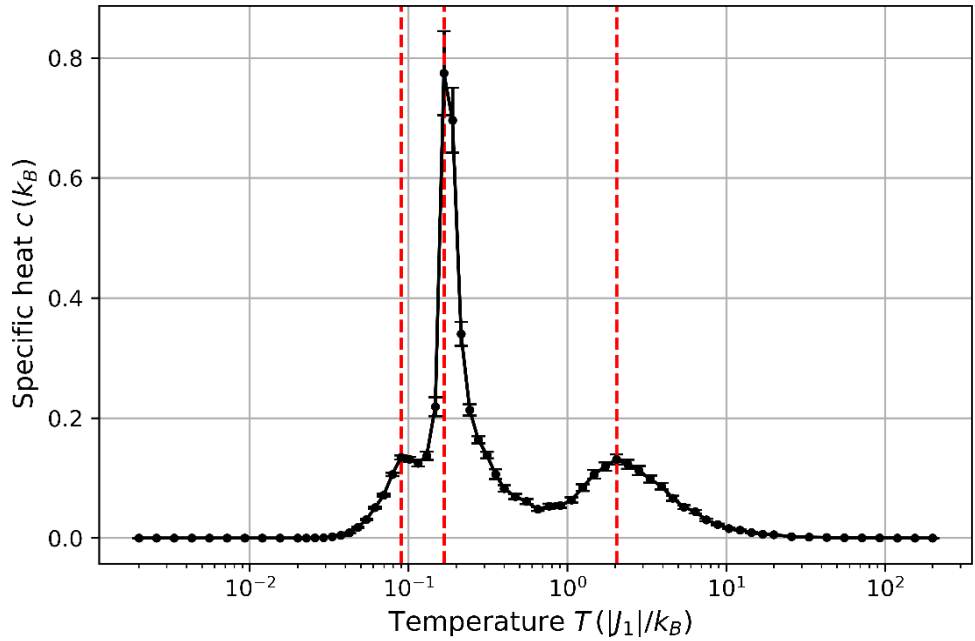

FIG. S1. Specific heat $c$ of the dipolar kagome spin ice model as a function of temperature $T$, obtained from Monte Carlo simulations. Dashed red lines indicate the curve's peaks, which occur at $k_B T/|J_1| = 2.0$, $0.17$ and $0.089$, defining the PM/SI1, SI1/SI2 and SI2/LRO phase boundaries, respectively. $J_1$ is the nearest neighbor coupling strength.

## 2. DISTRIBUTION OF LRO FRACTION IN MIXED SI2/LRO MICROSTATES

Our phase recognition algorithm requires, for each phase, a training set with a minimum number of data to ensure a given statistical confidence on its predictions. This is not a problem for the PM, SI1 and SI2 phases since they are extensively degenerate and, therefore, have a huge number of representative microstates. Thus, a very small fraction of them is enough to generate a training dataset. However, the ground state of DKSI is only sixfold degenerate, making it impossible to include the true LRO phase in our statistical analysis. To circumvent this issue, we exploited the fact that our simulated $12 \times 12$ kagome lattice is still able to fluctuate, due to finite size effects, for temperatures not too well below its SI2/LRO transition temperature ($\sim 0.089 \, |J_1|/k_B$), although such deviations from GS become more and more improbable as temperature decreases. Thus, we ran Monte Carlo simulations for five temperatures in the GS temperature range and, for each temperature, we waited long enough for the system to access new 20 distinct configurations. As a result, we collected 100 distinct microstates that are actually a mix of SI2 and LRO phases: most of the spins collectively follow the LRO pattern and the remaining part obeys only SI2 conditions (i.e., ice rules and magnetic charge ordering). Fig. S2 shows how the fraction of LRO distribute over these 100 microstates, varying between 0.84 and 0.89.

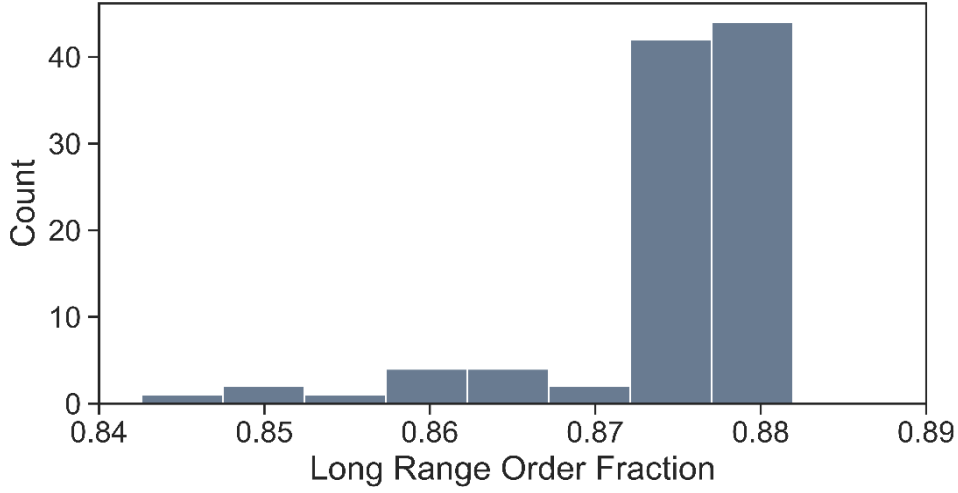

FIG. S2. Histogram of the fraction of LRO occuring in the 100 microstates sampled at the GS temperature range.

## 3. COMPUTATION OF THE BOUNDARY AND GROWTH FIELDS

The first and second boundary fields were identified, respectively, as the lowest and greatest fields at which the initial magnetization curve's first derivative suffers a sudden change. To calculate these values, we applied the Savitzky-Golay filter, with 5 points in the filter window and fitting by polynomial of order 3, to smooth the curves and obtain their derivatives. Fig. S3 illustrates this procedure for a particular SI2 curve (the same of Fig. 3 in the main text). One sees that there is an initial and a final interval in which the derivative has low and approximately constant values. To precisely define these boundaries, we considered a threshold value $C\sigma + \mu$, where $C$ is an arbitrary constant and $\mu$ and $\sigma$ are, respectively, the mean and standard deviation of the first derivative in the whole field interval. Choosing $C = 0.02$, we used a logic filter to evaluate when the derivative is above or below the threshold. This was described by the function

$$f(\mu_0 H) = \begin{cases} +1, & \text{if } \frac{d(M_y/M_s)}{d(\mu_0 H)} > C\sigma + \mu \\ 0, & \text{if } \frac{d(M_y/M_s)}{d(\mu_0 H)} = C\sigma + \mu \\ -1, & \text{if } \frac{d(M_y/M_s)}{d(\mu_0 H)} < C\sigma + \mu \end{cases}$$

and the result is shown in Fig. S3(b). The lowest and greatest field values for which $f(\mu_0 H) \neq -1$ were taken as $\mu_0 H_{B1}$ and $\mu_0 H_{B2}$, respectively.

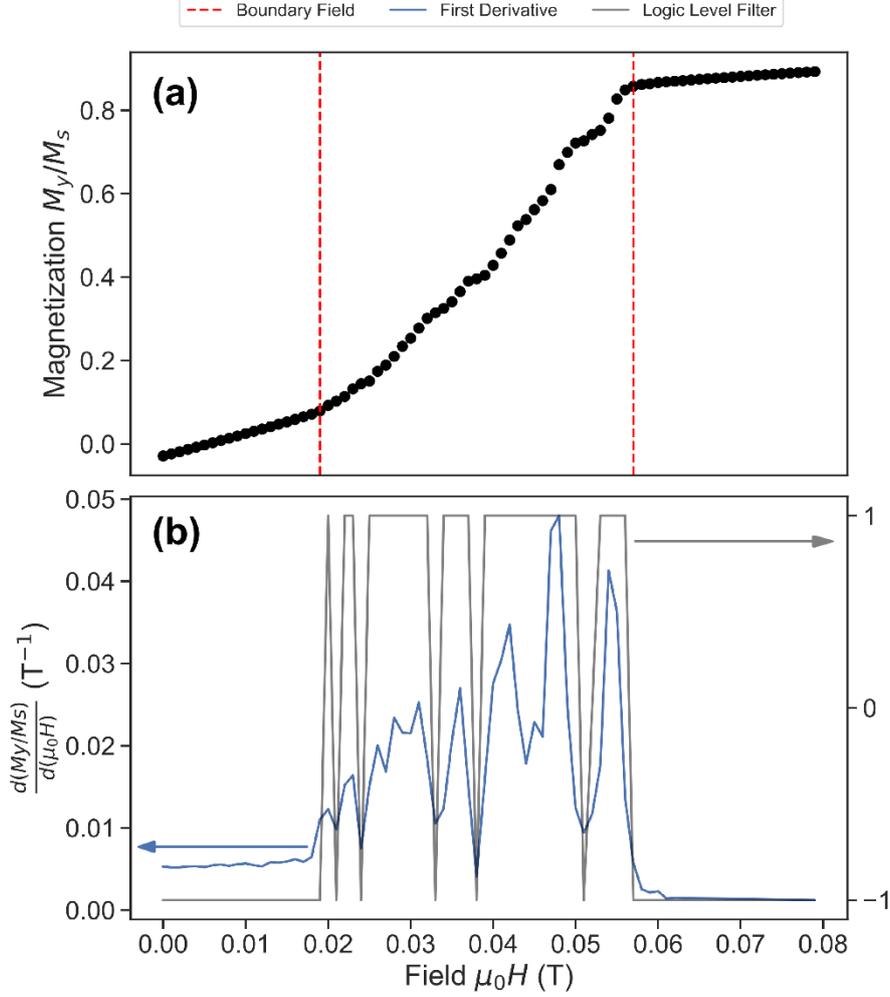

FIG. S3. (a) Initial magnetization curve (black dots) starting from a SI2 microstate and its boundary fields (red dashed lines). (b) The first derivative of the curve (blue) and the logic filter (black), which evaluates to +1 or -1 depending whether the derivative is above or below the threshold value, respectively.

Similar procedure was used to compute the growth field $\mu_0 H_G$, but restricted to the growth interval ($H_{B1} \leq H \leq H_{B2}$), as shown in Fig. S4. The Savitzky-Golay filter was applied with 5 points in the filter window using a fitted polynomial of order 3. Choosing $C = 1.1$, $\mu_0 H_G$ was taken as the lowest field for which $f(\mu_0 H) \neq -1$.

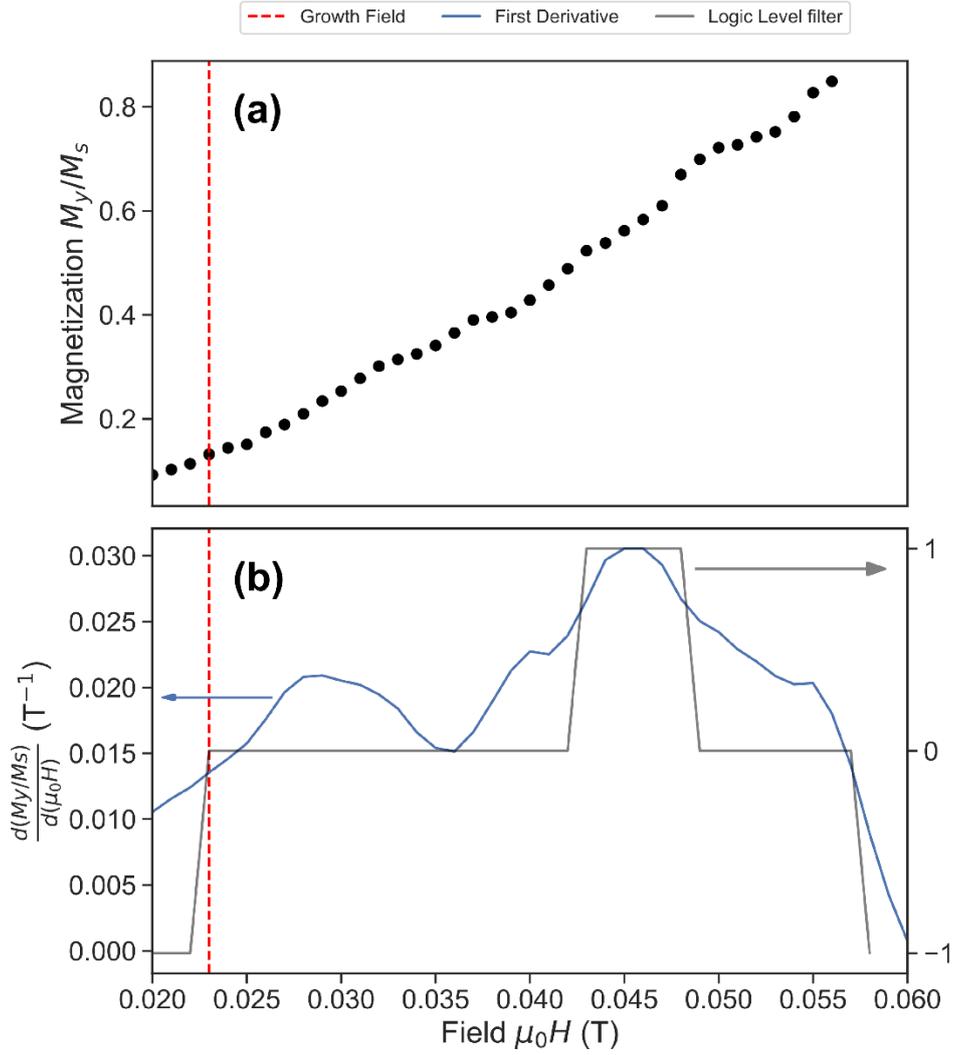

FIG. S4. (a) Growth interval of an initial magnetization curve (black dots) starting from a SI2 microstate and its growth field (red dashed line). (b) The first derivative of the curve (blue) and the logic filter (black), which evaluates to +1 or -1 depending whether the derivative is above or below the threshold value, respectively.

## 4. SECOND BOUNDARY FIELD AS A FUNCTION OF TEMPERATURE

Fig. S5 shows the second boundary field $H_{B2}$ as a function of effective temperature $T$. As noted in the main text, it remains essentially constant.

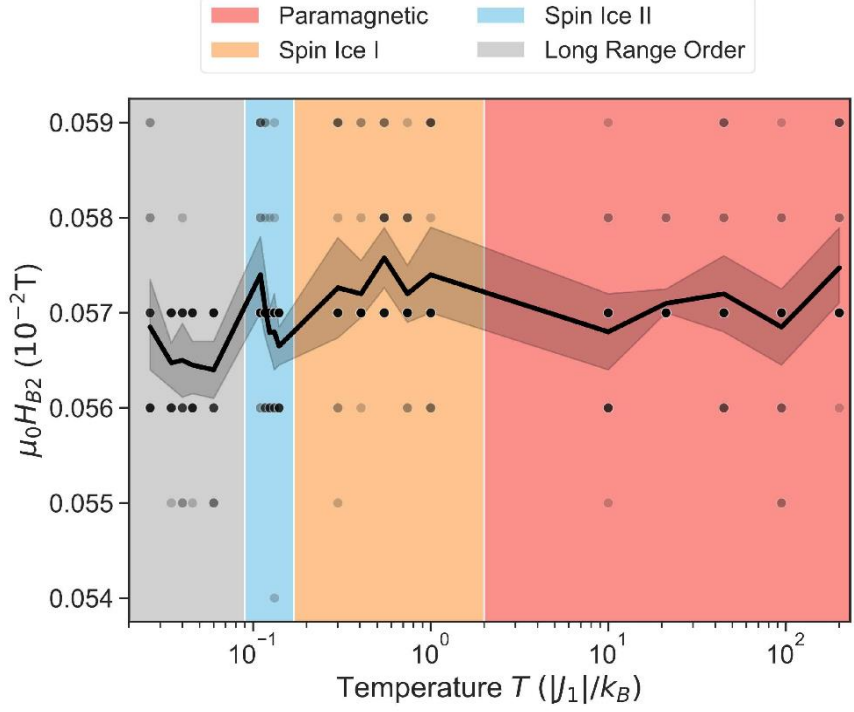

FIG. S5. Second boundary field $H_{B2}$ as a function of effective temperature $T$. Gray dots are the values for each sampled microstate and the black curve is the average over all microstates sampled at the same temperature.